\documentstyle[spie,psfig]{article}

\def\farcs{\hbox{$.\!\!^{\prime\prime}$}}

\title{A novel image reconstruction method applied to deep Hubble Space Telescope images} 

\author{A. S. Fruchter\supit{a} and R. N. Hook\supit{b} 
\skiplinehalf 
\supit{a}Space Telescope Science Institute\skipline 
3700 San Martin Drive \skipline
Baltimore, MD \hspace{0.5em}21210 \hspace{0.5em}USA 
\skiplinehalf 
\supit{b}Space Telescope European Coordinating Facility\skipline
D-85748 Garching \hspace{0.5em} Germany
}

\authorinfo{The authors can be reached electronically at fruchter@stsci.edu and
rhook@eso.org.}

\begin{document} 
  \maketitle 

%%%%%%%%%%%%%%%%%%%%%%%%%%%%%%%%%%%%%%%%%%%%%%%%%%%%%%%%%%%%% 
\begin{abstract}
We have developed a method for the linear reconstruction of an image
from undersampled, dithered data, which has been used to create the
distributed, combined Hubble Deep Field\cite{hdf+96} images -- the deepest optical
images yet taken of the universe.  The algorithm, known
as Variable-Pixel Linear Reconstruction (or informally as
``drizzling"), preserves photometry and resolution, can weight input
images according to the statistical significance of each pixel, and
removes the effects of geometric distortion both on image shape and
photometry.  In this paper, the algorithm and its implementation are
described, and measurements of the photometric accuracy and image
fidelity are presented.  In addition, we describe the
use of drizzling to combine dithered images in the presence of cosmic
rays.
\end{abstract}

%>>>> Please include a list of keywords right after the abstract 

\keywords{image reconstruction, image restoration, undersampled images, Hubble Space Telescope}

%%%%%%%%%%%%%%%%%%%%%%%%%%%%%%%%%%%%%%%%%%%%%%%%%%%%%%%%%%%%%
\section{INTRODUCTION} 

The Hubble Space Telescope (HST) is now capable of providing the superb
images for which it was designed.  However, the detectors
on HST are only able to take full advantage of the full resolving power
of the telescope over a limited field of view.  
In particular, the primary optical imaging camera on the HST, the Wide
Field and Planetary Camera 2\cite{tbbc+94}, is composed of four separate
800x800 pixel CCD cameras, one of which, the planetary
camera (PC) has a scale of $0\farcs046$ per pixel, while the other three,
arranged in a chevron around the PC, have a scale of $0\farcs097$ per
pixel.  These latter three cameras, referred to as the wide field
cameras (WFs), are the primary workhorse for deep imaging surveys on HST.
However, these cameras greatly undersample the HST image.  The width
of a WF pixel
equals the full-width a half maximum  of the optics in the the 
near-infrared, and greatly exceeds
it in the blue.   The effect of undersampling
on WF images is illustrated by the "Great Eye Chart in the
Sky" in Figure 1.  

\begin{figure}
\centerline{\psfig{figure=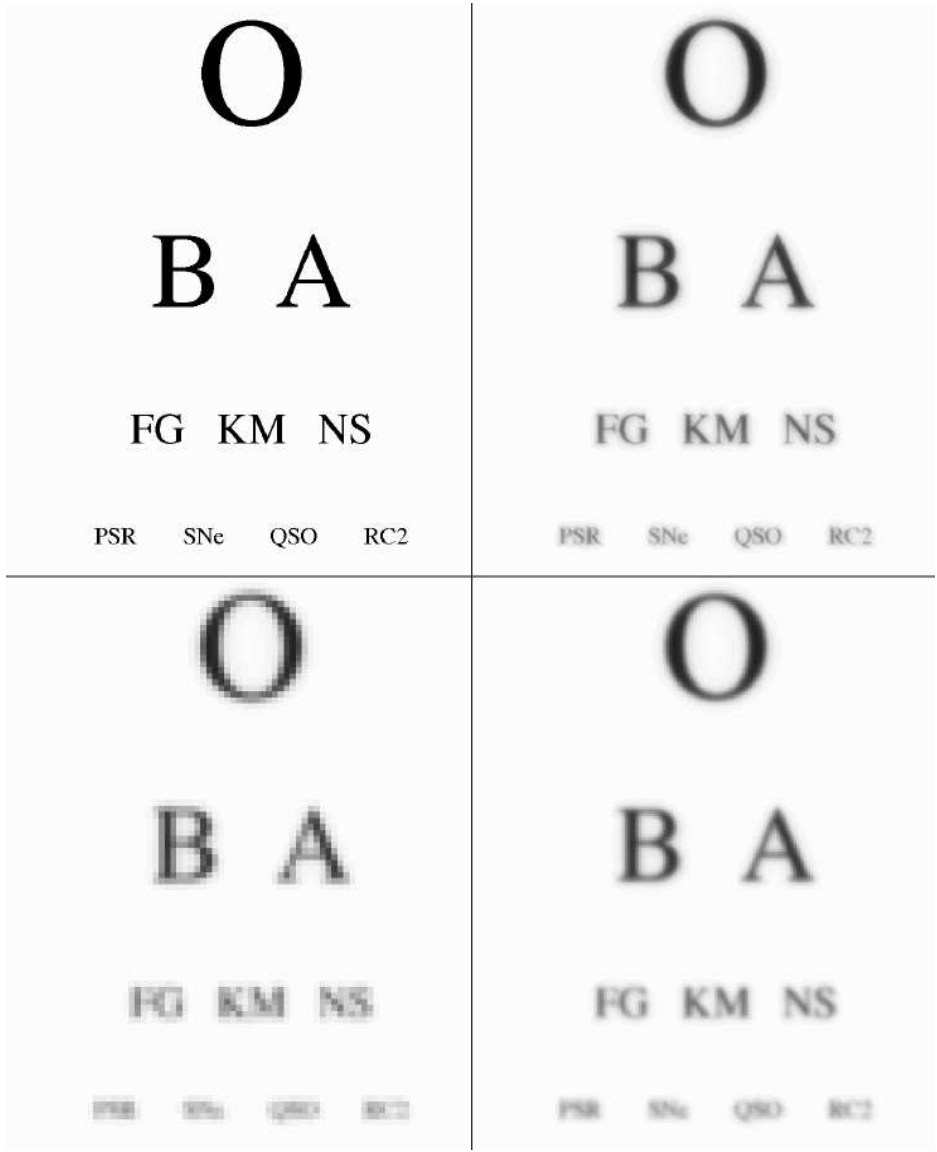,height=6.0in}}
\caption{In the upper left corner of this figure, we present the ``true image",
{\it i.e.} the image one would see with an infinitely large telescope.  The
upper right shows the image after convolution with the optics of the
Hubble Space Telescope and the
WFPC2 camera -- the primary wide-field imaging instrument presently 
installed on the HST.  The lower
left shows the image after sampling by the WFPC2 CCD, and the lower right
shows a linear reconstruction of dithered CCD images.
\label{fig:bigeye}}
\end{figure}

Fortunately, much of the information lost to undersampling
can be restored.  In the lower right of Figure 1
we display a restoration made using one of the family of techniques we refer
to as ``linear reconstruction."  The most commonly used of these techniques are
shift-and-add and interlacing.
The image in the lower right corner has been restored by
interlacing dithered images.  
However, due to the occasional small positioning errors of telescope and the 
non-uniform shifts in pixel space caused by the 
geometric distortion of the optics, true interlacing of
 HST images is
often infeasible.  The other standard linear reconstruction technique,
shift-and-add can easily handle arbitrary dither postions, but it
convolves the image yet again with the orginal pixel, adding to the
blurring of the
image and the correlation of the noise.
The importance of avoiding unnecessarily convolving the image
with the pixel is emphasized by comparing the upper and lower
right
hand images in Figure 1.
The deterioration in image quality
is due entirely to convolution of the image by the WF pixel. 
Here we present a new method
which has the versatility of shift-and-add yet largely maintains
the resolution and independent noise statistics of interlacing.

%%%%%%%%%%%%%%%%%%%%%%%%%%%%%%%%%%%%%%%%%%%%%%%%%%%%%%%%%%%%%
  \section{THE METHOD}

The Drizzle algorithm is conceptually straightforward. Pixels
in the original input images
are mapped into pixels in the subsampled output image, taking into account
shifts and rotations between images and the optical distortion of the camera.
However, in order to avoid convolving the image with the large pixel "footprint"
of the camera, we allow the user to shrink the pixel before it is averaged
into the output image, as shown in Figure 2.

\begin{figure}

\centerline{\psfig{figure=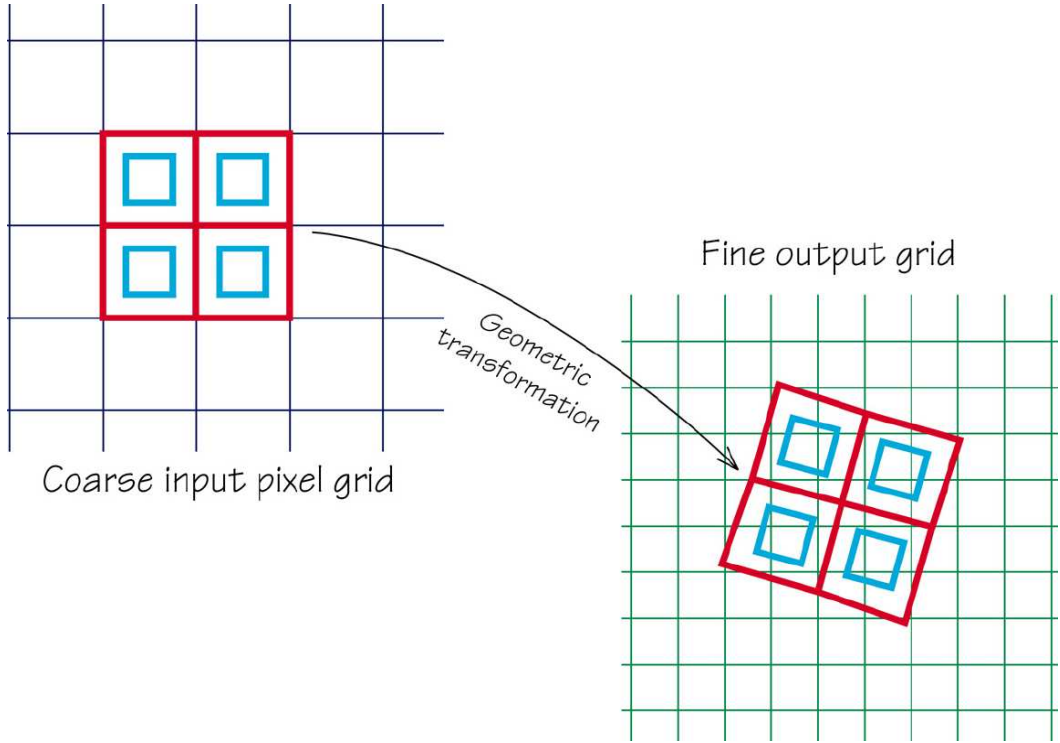,width=5.5in}}
\caption{A schematic representation of Drizzling.  The input pixel grid 
(shown on the left) is mapped onto a finer output grid (shown on right),
taking into accounts shift, rotation and geometric distortion.  The user
is allowed to "shrink" the input pixels to smaller pixels, which we refer to
as drops (faint inner squares).   A given input image only affects 
output image pixels under drops.  In this particular case, the central
output pixel receives no information from the input image.} 
\end{figure}

The new shrunken pixels, or ``drops",  rain
down upon the subsampled output
In the case of the HDF, the drops used had
linear dimensions one-half that of the input pixel --- slightly larger than
the dimensions of the output subsampled pixels. 
The value of an input pixel is averaged into an output pixel with 
a weight proportional to the area of overlap between the ``drop" and
the output pixel.
Note that if the drop size is sufficiently small
not all output pixels have data added to them from each
input image.   One must therefore choose a drop size that is small
enough to avoid degrading the image, but large enough so that 
after all images are ``dripped" the coverage is fairly uniform.

The drop size is controlled by a user-adjustable parameter called
{\bf pixfrac}, which is simply the ratio of the linear size of the drop
to the input pixel (before any adjustment due to the geometric distortion
of the camera).   Thus interlacing is equivalent to setting 
${\rm{\bf pixfrac}}=0.0$,
while shift-and-add is equivalent to ${\rm{\bf pixfrac}}=1.0$.

When a drop with value
$i_{xy}$ and user defined weight $w_{xy}$ is added
to an image with pixel value $I_{xy}$, weight $W_{xy}$,
and fractional pixel overlap $0 < a_{xy} < 1$, the resulting
value of the image $I'_{xy}$
and weight $W'_{xy}$ is

\begin{eqnarray}
W'_{xy}  &=& a_{xy}w_{xy} + W_{xy} \\
I'_{xy} &=& \frac{a_{xy}i_{xy}w_{xy} + I_{xy}W_{xy}}{W'_{xy}}
\end{eqnarray}

This algorithm has a number of advantages over the more standard
linear reconstruction methods presently used.
Since the area of the pixels scales with the Jacobian of the 
geometric distortion, this algorithm preserves both surface and 
absolute photometry.   Therefore
flux can be measured using an aperture whose size is
independent of position on the chip.  
As the method anticipates that  a given output pixel may receive
no information from a given input pixel, missing data (due for
instance to cosmic rays or detector defects) does not cause
a substantial problem, so long as there are enough dithered
images to fill in the gaps caused by these zero-weight input pixels.
Finally, the linear weighting scheme is
statistically optimum when inverse variance maps are used as weights.

%These weights may vary spatially to accommodate changing signal-to-noise
%ratios across input frames (e.g. due to variable scattered light). The
%final output weighting image (an inverse variance map) is saved as
%well as the combined image frame and can be used in further analysis.

\section{Image Fidelity}
The drizzling algorithm was designed to obtain optimal signal-to-noise
on faint objects while preserving image resolution.
These goals are unfortunately not fully compatible.  For example non-linear
image restoration procedures which attempt to remove the blurring of
the PSF and the pixel by enhancing the high frequencies in the image (such
as such as the Richardson-Lucy\cite{ric72,luc74,lh91}
and maximum entropy methods\cite{gd78,wd90}) directly exchange signal-to-noise for resolution.
In the drizzling algorithm no compromises on signal-to-noise have been made; 
the weight of an input pixel in the final output image is entirely
independent of its position on the chip.   Therefore, if the dithered images
do not uniformly sample the
field, the ``center of light'' in an output pixel may be offset from the center
of the pixel, and that offset may vary between adjacent pixels.  
The large dithering offsets which may
be used in WFPC2 imaging combined
with geometric distortion can 
produce a sampling pattern that varies across the
field.
The output PSFs produced by the combination of such irregularly
dithered datasets may, on occasion, show significant
variations about the best fit Gaussian. 
Fortunately this
does not noticeably affect aperture photometry performed with typical
aperture sizes.    In practice the variability about the
Gaussian appears larger in WFPC2 data
than our simulations would lead us to
expect.  Examination of recent dithered stellar fields
leads us to suspect that this excess variability results from a problem with the
original data, possibly caused by charge transfer errors in the CCD.

\section{Photometry}
 
The WFPC2 optics geometrically distort the images: pixels at the corner
of each CCD subtend less area on the sky than those near the center.
However, after application of the flat field, a source of uniform surface
brightness on the sky  produces uniform counts across the CCD.  Therefore
point sources near the corners of the chip are artificially brightened
compared to those in the center.
 
In order to study the ability of Drizzle to remove the photometric 
effects of geometric distortion, we have created a 
a sub-sampled grid of
$19 \times 19$ artificial stellar PSFs and adjusted their counts to
reflect the effect of geometric distortion --  the stars
in the corners are up to $ \sim 4 \%$ brighter than those in the center
of the chip. This image was then shifted and down-sampled onto four simulated
WF frames 
and the results combined using drizzling with typical parameters.
Aperture photometry on the $19 \times 19$ grid after drizzling reveals that
the effect of geometric distortion on the photometry has been dramatically
reduced: the RMS photometric variation in the drizzled image is 0.004 mags.
 
\section{Cosmic Ray Detection}
 
Few HST observing proposals have sufficient time to take a number of
exposures at each of several dither positions.
Therefore, if dithering is to be of wide-spread use, one must be
able to remove cosmic rays from data where few, if any,
images are taken at the same position on the sky. We have therefore
begun to adapt Drizzle to the removal of cosmic rays.

A technique which appears quite
promising is described below:
 
\begin{enumerate}
\item{Drizzle each image onto a separate sub-sampled output image using 
$pixfrac = 1.0$}
 
\item{Take the median of the output drizzled images.}
 
\item{Map the median image back to the input plane of each of the 
individual images, taking into accout the
image shifts and geometric distortion.  The is now done by interpolating
the values of the median image using a program
we call ``Blot''.}  
 
\item{Take the spatial derivative of each of the blotted output images.}
 
\item{Compare each original image with the corresponding blotted image.  Where th
e difference
is larger than can be explained by noise statistics, or the flattening effect of
 taking the median
or perhaps an error in the shift
(the magnitudes of the latter two effects are estimated using the image 
derivative), the suspect pixel is masked.}
 
\item{Repeat the previous step on pixels adjacent to pixels already masked, using
 a more stringent comparison.}

\item{Finally, drizzle the input images onto a single output image
using the pixel masks created in the previous steps.}

\end{enumerate}

Figure 3 shows the result of applying this method to data originally
taken by Cowie and colleagues\cite{chs95}, which we have reprocessed
using Drizzle.

\section{The Hubble Deep Field Images}

Drizzle was originally developed for the Hubble Deep Field, 
a project whose purpose was to image an otherwise unexceptional region
of the sky to depths far beyond those of previous astronomical images.
Exposures were taken in four color bands from the near ultraviolet to the
near infra-red.  The resulting images are available in the published
astronomical literature\cite{hdf+96} as well as from the Space Telescope
Science Institute via the World Wide Web at 
http://www.stsci.edu/ftp/science/hdf/hdf.html.

\begin{figure}[t]

\centerline{\psfig{figure=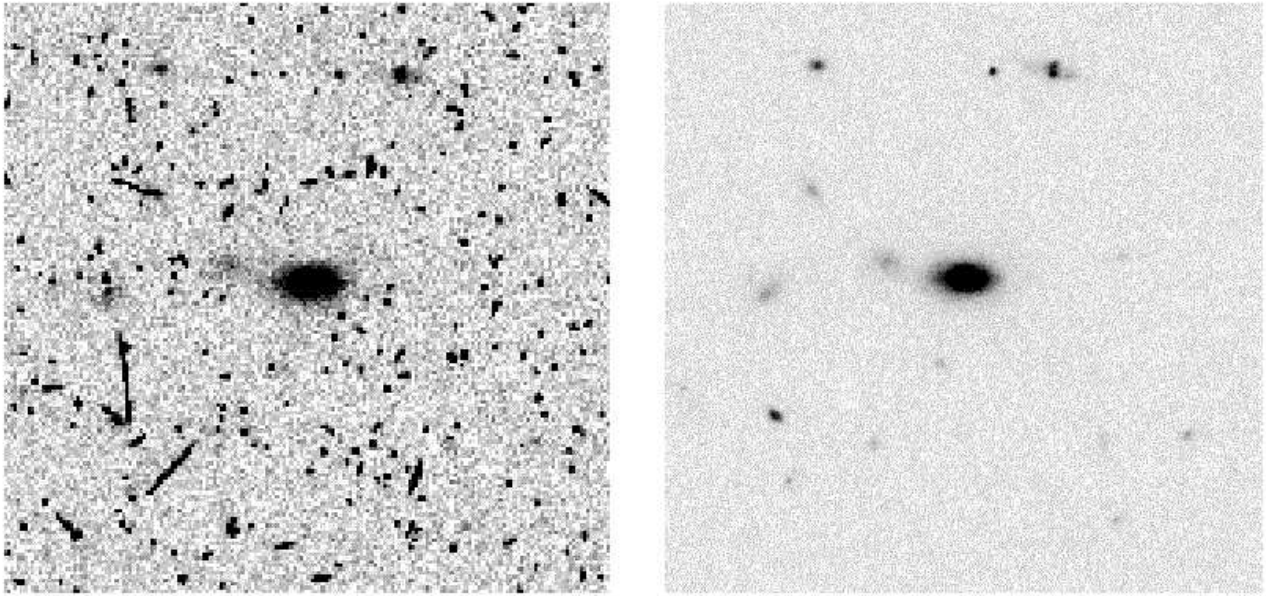,width=6.75in}}

\caption{The image on the left shows a region
of one of twelve 2400s archival images taken 
with a wide near-infrared filter on WFPC2.
Numerous cosmic
rays are visible.  On the right is the drizzled combination of the twelve
images, no two of which shared a dither position.}

\end{figure}

%%-----------------------------------------------------------

%%%%% References %%%%%

  \bibliography{journals,mine,drizzle}   %>>>> bibliography data in report.bib
  \bibliographystyle{spiebib}   %>>>> makes bibtex use spiebib.bst
 
  \end{document}